\documentclass[a4paper,onecolumn,nofootinbib]{revtex4}
\usepackage{amssymb}
\usepackage{latexsym}
\usepackage{epsfig}
\usepackage{amsmath}
\usepackage{graphicx}
\usepackage{mathtools}
\usepackage{bm}
\usepackage{cancel}
\usepackage{amssymb}
\usepackage{float}
\usepackage{subfigure}
\usepackage{dcolumn}
\usepackage[T1]{fontenc}
\usepackage[utf8]{inputenc}
\usepackage[colorlinks]{hyperref}
\usepackage[usenames,dvipsnames]{color}

\hypersetup{
     breaklinks=true,
    pdfstartview={FitH},    
    colorlinks=true,       
    linkcolor=blue,          
    citecolor=red,        
    filecolor=magenta,      
    urlcolor=blue,           
    anchorcolor=green,      
    linktocpage=true
}
\begin{document}
\title{The Lambert $W$ equation of state in light of DESI BAO}
\author{ Vipin Chandra Dubey}
\email{vipindubey476@gmail.com} 
\affiliation{Department of Mathematics, Institute of Applied Sciences and Humanities, GLA University, Mathura 281406, Uttar Pradesh, India}
\author{Subhajit Saha}
\email{subhajit1729@gmail.com}
\affiliation{Department of Mathematics, Panihati Mahavidyalaya, Kolkata 700118, West Bengal, India}
\author{Abdulla Al Mamon\footnote{Correspondiong author}}
\email{abdulla.physics@gmail.com}
\affiliation{Department of Physics, Vivekananda Satavarshiki Mahavidyalaya (affiliated
to the Vidyasagar University), Manikpara-721513, West Bengal, India}

\begin{abstract}
We investigate a unified dark-fluid model whose effective equation of state (EoS) is described by logarithmic and power-law terms involving the Lambert $W$ function. The model parameters are constrained using BAO data, including DESI measurements, together with Pantheon+ Type Ia supernova observations and direct Hubble parameter measurements. The analysis yields $\theta_{1}=0.087\pm 0.011$, $\theta_{2}=-3.35\pm 0.13$, $r_d = 146\pm 2.5$~Mpc, and $H_0 = 67.4 \pm 1.2~\text{km\,s}^{-1}\text{Mpc}^{-1}$. The inferred Hubble constant, $H_{0}$, is consistent with the Planck 2018 measurement and remains in tension with local determinations, thereby reflecting the Hubble tension. We further examine the evolution of deceleration, effective EoS, and jerk parameters, complemented by the $Om(z)$ diagnostic. Our analysis reveals that the model provides a consistent description of late-time cosmic acceleration. Finally, the observational viability of the model is assessed using Akaike and Bayesian information criteria and compared with that of the standard $\Lambda$CDM model.
\end{abstract}
\maketitle

\section{Introduction}
Observational evidence~\cite{acc1,acc2,acc3,acc4,acc5} indicates that the Universe is currently undergoing an accelerated expansion. Although numerous cosmological models have been proposed, none of them has achieved full agreement with all astronomical observations. Among these, the $\Lambda$ cold dark matter ($\Lambda$CDM) model stands out, demonstrating consistently strong agreement with a wide range of observational data sets. In this framework, a tiny cosmological constant, $\Lambda$, acts as \emph{dark energy} --- the dominant contributor to the energy density of the Universe, whereas \emph{cold dark matter} behaves as pressureless dust. When combined, these two components account for roughly 96\% of the total energy budget of the Universe. However, it faces fundamental theoretical challenges associated with the cosmological constant, including the \emph{fine-tuning} and \emph{cosmic coincidence} problems \cite{refft1,refccp1}. Moreover, increasingly precise observations have revealed significant tensions, most notably the $H_0$ and $S_8$ tensions, which challenge the completeness of the concordance $\Lambda$CDM model \cite{refs8,refh01,refh02,refh03}. The $H_0$ tension arises from the disagreement between the value of the Hubble constant inferred from Planck cosmic microwave background (CMB) data within the $\Lambda$CDM framework, $H_0 = 67.36 \pm 0.54~\mathrm{km\,s^{-1}\,Mpc^{-1}}$ \cite{acc5}, and the significantly higher value measured locally by the SH0ES collaboration, $H_0 = 73.04 \pm 1.04~\mathrm{km\,s^{-1}\,Mpc^{-1}}$ \cite{refh02}. This discrepancy now exceeds the $5\sigma$ level, and its persistence suggests the need for alternative models beyond the standard $\Lambda$CDM model. Moreover, simulations generated by the $\Lambda$CDM model have predicted certain features on the galactic and sub-galactic scales that are not consistent with observations. These problems are collectively known as \emph{``small-scale-Cosmology crisis"} and include the core-cusp problem, the missing satellite or dwarf galaxy problem, and the ``too big to fail" problem \cite{Mavromatos1}.\\

The weaknesses of the $\Lambda$CDM paradigm led cosmologists to look for alternative and more simplified models. A natural question to ask was whether single dark fluid models could provide a theoretical framework for unifying dark matter (DM) and dark energy (DE). Indeed, such unification has two notable advantages. Firstly, only a single component is required to explain both the observed late-time accelerated expansion and the formation of structures in the Universe, and secondly, it allows us to treat DM and DE in the same way at the perturbation level \cite{Hu1}. Interestingly, the Chaplygin gas (CG) model \cite{refum1,refum2}, as well as its generalizations~\cite{refum3,Sen1,refum4,refomaam,refum5} were proposed, which provide such a unified framework. The CG-class of models achieve this unification by considering equations of state that are polytropic in nature. Remarkably, the CG behaves like pressureless, dust-like matter in the early Universe, and at late times, it evolves to behave like a cosmological constant. Despite these elegant features, the CG models have been shown to be unstable at the perturbative level \cite{Fabris1} and, more importantly, they do not agree well with observational data, particularly with the CMB anisotropy measurements~\cite{refdbum1,refdbum2}. Alternative single dark fluid models that describe the smooth transition of the Universe from a decelerating phase to the late-time acceleration phase are the logotropic DE model and its generalizations, studied extensively in the literature \cite{Chavanis1,Chavanis2,Ferreira1,Chavanis3,Chavanis4,Boshkayev0,Mamon0,Mandal:2021acs,Benaoum1}, an ideal gas with a varying negative absolute temperature \cite{Saha0}, and fluids obeying the Anton-Schmidt's equation of state within the Debye approximation, studied by Capozziello and others \cite{Capozziello1,Capozziello2}. Note that the logotropic DE fluid is a limiting case of the Anton-Schmidt fluid. Statistical analysis by Boshkayev et al. \cite{Boshkayev1} revealed that the $\Lambda$CDM paradigm is statistically better compared to the logotropic and generalized logotropic models. In fact, pure logotropic models are incapable of explaining the dynamics of the Universe. Very recently, the impact of most of these models has been investigated in light of the Dark Energy Spectroscopic Instrument (DESI) Data Release 1 (DR1) data \cite{Carloni1}. Surprisingly, the study showed evidence for a log-corrected form of DE, although this result is in tension with that obtained in Ref. \cite{Boshkayev1}.\\

An extensive study of the existing literature on single dark fluid models shows that the superiority of these models is not established beyond reasonable doubt. So, there still exists sufficient opportunity for cosmologists to explore new single dark fluid models, even if they are phenomenological or seem rather ad hoc. Motivated by these observations, Saha and Bamba \cite{Saha1} proposed an effective EoS which is a sum of a logarithmic term and a power-law term. An interesting feature of this EoS is that both terms are associated with the Lambert $W$ function. The evolution of the deceleration parameter in the model shows that this single dark fluid can explain the smooth transition of the decelerated Universe into the late-time acceleration phase. This model has been analyzed under observational constraints using Hubble parameter data only, and various model parameters were studied by Mamon and Saha \cite{Mamon1}. Subsequently, Banerjee et al. \cite{Banerjee1} extended the investigation to the perturbative regime. Their findings show that introducing a Lambert $W$ -type dynamical dark energy component leads to notable modifications in the growth rate of cosmic structures and produces significant effects on matter density fluctuations in the Universe. In this work, we undertake a comprehensive Bayesian analysis of the Lambert $W$ dark fluid model in the context of a combined observational data, including Pantheon$+$ compilation of Type Ia supernova data, BAO measurements from DESI DR1, and direct measurements of the Hubble parameter. The paper is organized as follows. In Section \ref{sec-mode}, we present a concise description of the Lambert $W$ dark fluid model. A Bayesian analysis of the model is described in the next two sections, with the data and methodology outlined in Section \ref{sec-data}, and a detailed analysis presented in Section \ref{sec-results}. Finally, a summary of the work and our deductions are provided in Section \ref{sec-cons}.

\section{The model with the Lambert $W$ equation of state}\label{sec-mode}
The Lambert $W$ function is a special function in Mathematics which has found useful applications in various branches of science and engineering; cf. \cite{Veberic1} and the references therein. The first mention of the mathematical problem involving this special function is attributed to Euler \cite{Euler1}, although he credited Lambert for the latter's earlier work \cite{Lambert1} on the subject.\\

The Lambert $W$ function, also known as the ``product logarithm”, is defined as the solution of the transcendental equation
\begin{equation}\label{eqtransapplica}
xe^x=k,
\end{equation}
i.e.,
\begin{equation} \label{lw-1}
\text{Lambert}~W(k) \cdot e^{\text{Lambert}~W(k)}=k.
\end{equation} The above equation admits two real solutions if $-\frac{1}{e}<k<0$, which correspond to two real branches of Lambert $W$ \cite{Veberic2}. It is possible to obtain many solutions for complex values of $k$, which correspond to infinitely many complex branches of Lambert $W$ \cite{Corless1,Jeffrey1}. Some important values of the Lambert \(W\) function are obtained at \(k=-\frac{1}{e}\), \(k=0\), and \(k=1\), yielding Lambert $W(-\frac{1}{e})=-1$, Lambert $W(0)=0$, and Lambert $W(1)\approx 0.567143$. The value Lambert $W(1) \approx 0.567143$ is referred to as the \emph{omega constant}. This notable constant emerges naturally in exponential equations and is often viewed as an exponential analogue of the golden ratio because of its distinctive fixed-point property and its frequent occurrence in self-referential exponential relations. One might refer to the book by Mez\~{o} \cite{Mezo1} for an extensive and insightful discussion on the properties and applications of the Lambert $W$ function. In the following, the Lambert $W$ function will be denoted simply by $W$. For a historical account of the notation $W$, see Hayes \cite{Hayes1}. The $N^{\text{th}}$ derivative of $W$  can then be expressed as follows: 
\begin{equation}
W^{N}(k)=\frac{W^{N-1}(k)}{k^N[1+W(k)]^{2N-1}}\varphi _{y=1}^{N} \delta _{yN}W^{y}(k),~~~~k \neq -\frac{1}{e},
\end{equation}
where $\delta _{yN}$ denotes the coefficients of the following number triangle:
\[ \begin{array}{ccccc}
\phantom{+}1 & & & &\\
-2 & -1 & & & \\
\phantom{+}9 & \phantom{+}8 & \phantom{+}2 & & \\
-64 & -79 & -36 & -6 & \\
\phantom{+}625 & \phantom{+}974 & \phantom{+}622 & \phantom{+}192 & \phantom{+}24  
\end{array} .\]
As a special case, the first-order ($N=1$) derivative of $W(k)$ can be evaluated as 
\begin{eqnarray}
W'(k) &=& \frac{W(k)}{k[1+W(k)]}, ~\mbox{if~} k \neq 0 \nonumber \\
&=& \frac{e^{-W(k)}}{1+W(k)}.
\end{eqnarray}
Similarly, the antiderivative of $W(k)$ is given by
\begin{equation}
\int W(k)\text{d}k=k\left[W(k)+\frac{1}{W(k)}-1\right]+{\cal{C}}^{\prime},
\end{equation}
where ${\cal{C}}^{\prime}$ denotes the constant of integration.\\

It is worth mentioning that the inverses of the Moyal function and the Gaisser-Hillas function, both of which are intimately connected to the Lambert $W$ function, have useful applications in the field of Astrophysics \cite{Veberic1}. In Cosmology, Thompson \cite{Thompson1,Thompson2,Thompson3} has recently studied the implications of this function in the LHQ model, an alternative theory of Cosmology. This model is unique in the sense that it allows for the possibility that the Universe existed before the Big Bang. In fact, this model divides the timeline of the Universe into two distinct epochs --- the pre-Big Bang epoch and the post-Big Bang epoch. It admits a scalar field described by a simple expression involving the Lambert $W$ function, and the pre-Big Bang era, extending from the infinite past up to the Big Bang singularity, occupies a vast portion on the principal branch compared to the canonical post-Big Bang era. It is interesting to note that the Big Bang is a natural manifestation of the contraction of the pre-Big Bang Universe to the Big Bang singularity in this model \cite{Thompson3}.\\

Prior to Thomson's work, Saha and Bamba \cite{Saha1} introduced an (effective) equation of state of the cosmic fluid that involved the Lambert $W$ function. Although the study was quite rudimentary, it demonstrated that this equation of state could, in principle, provide a unified description of DM and DE. In this regard, one must note that the Lambert $W$ function appears implicitly when deriving solutions of the energy-momentum conservation equation in the gravitationally induced adiabatic matter creation scenario \cite{Chakraborty1} and this fact somewhat motivated the authors in \cite{Saha1} to study whether some suitable expression of the Lambert $W$ function is somehow connected with the evolution of the Universe.\\ 

Let us assume a spatially flat, homogeneous and isotropic Friedmann-Lema\^itre-Robertson-Walker (FLRW) universe governed by the metric\footnote{We consider relativistic units $G=1=c$.}
\begin{equation}
ds^2=-dt^2+a^2(t)\left[dr^2+r^2(d\theta^2+\mbox{sin}^2\theta d\phi^2)\right],
\end{equation}
where $a(t)$ is the scale factor of the Universe. On the basis of Weyl's postulate, we further assume that the Universe is filled with a perfect fluid having an energy-momentum (EM) tensor
\begin{equation}
T_{\mu\nu}=(\rho+p)u_{\mu}u_{\nu}+pg_{\mu\nu}, 
\end{equation}
where $\rho$ and $p$ are, respectively, the energy density and pressure of the cosmic fluid and $u^{\alpha}=(1,0,0,0)$ is the four-velocity of the fluid, normalized as $u^{\alpha}u_{\alpha}=-1$. The Einstein's field equations yield the Friedmann and the acceleration equations, given respectively, by
\begin{eqnarray}
H^2 &=& \frac{8\pi\rho}{3} \label{feq}\\
\dot{H} &=& -4\pi(\rho+p), \label{aeq}
\end{eqnarray}
where $H=\frac{\dot{a}}{a}$ is the Hubble parameter. We also have the EM conservation equation
\begin{equation}
\dot{\rho}+3H(\rho+p)=0. \label{ceq}
\end{equation}
We observe that Eqns. (\ref{feq}), (\ref{aeq}), and (\ref{ceq}) are not independent. In order to solve these equations, we must prescribe a relation between $\rho$ and $p$. This relation is known as the equation of state (EoS). The authors in \cite{Saha1} considered an (effective) EoS of the form
\begin{eqnarray}\label{eq-weffz}
\omega_{{\rm{eff}}} &=& \frac{p}{\rho} \nonumber \\
&=& \theta_1 \mbox{ln}\{W\left(a\right)\}+\theta_2 \{W(a)\}^3,
\end{eqnarray}
where the parameters $\theta_1$ and $\theta_2$ must be fixed from the observations. They estimated the values of these parameters to be $\theta_1=\frac{1}{7}$ and $\theta_2=-\frac{16}{5}$ using some useful initial conditions. However, in a subsequent work, Mamon and Saha \cite{Mamon1} obtained the best-fit values $\theta_1=-0.166_{-0.104}^{+0.104}$ and $\theta_2 = -4.746_{-0.479}^{+0.479}$ at the $1\sigma$ level. These values show modest deviations from the theoretically estimated values in \cite{Saha1}. This is most certainly due to the fact that the authors in \cite{Mamon1} considered only the Hubble parameter data to constrain these parameters.\\

The motivation for considering the EoS given in Eq. (\ref{eq-weffz}), which contains the Lambert $W$ function, originates from its rich nonlinear behavior and its ability to interpolate smoothly between different asymptotic regimes.  In a unified cosmological framework, a viable cosmic fluid must mimic pressureless matter at early times to support structure formation and subsequently develop negative pressure (i.e., long-range repulsive effects) to account for the observed late-time acceleration of the Universe. The Lambert $W$ functional form naturally provides such a transition through a single analytic expression. Furthermore, the Lambert $W$ function naturally emerges when thermodynamic variables are related through transcendental equations involving products of a variable and its exponential (as given in Eq. (\ref{eqtransapplica})). Such relations commonly appear in systems with nonlinear interactions, excluded-volume effects, quantum-statistical corrections, self-consistent mean-field treatments, and modified gravitational or cosmological frameworks. Inspired by these properties, Saha and Bamba \cite{Saha1} introduced a Lambert W-based equation of state for the dark sector, demonstrating that it can reproduce the main evolutionary phases of the Universe within a single-fluid framework, without introducing separate dark matter and dark energy components. Therefore, the principal motivation of the model is its capability to capture the transition from matter domination to accelerated expansion while retaining mathematical simplicity and analytical tractability.\\

As is well known, the deceleration parameter $q$ is a kinematical quantity that characterizes the rate at which the expansion of the universe changes. It is dimensionless and provides insight into whether the cosmic expansion is accelerating or decelerating over time. It is defined as:
\begin{equation}    
 q =-\frac{\ddot{a}}{aH^2} =-1- \frac{\dot{H}}{H^{2}} 
\end{equation}
where overdots denote derivatives with respect to cosmic time $t$, respectively. A negative $q$ indicates that the universe is accelerating, while a positive $q$ reflects a decelerating expansion. For this model, the deceleration parameter $q$ as a function of redshift $z = \frac{1}{a} - 1$ is calculated as \cite{Saha1}
\begin{eqnarray} \label{eq-qevo}   
 q(z) &=& -1+\frac{3}{2} (1+\omega_{\rm{eff}}) \nonumber \\ 
 &=& \frac{3}{2} \left[ 1 + \theta_1 \ln W\left(\frac{1}{1+z}\right) + \theta_2 W\left(\frac{1}{1+z}\right)^3 \right] - 1.
\end{eqnarray}

To further investigate the evolution of cosmic acceleration, we consider an additional key kinematical quantity, the jerk parameter~\cite{refj1,refj2,refj3,refzt5}, which depends on the third order derivative of the scale factor $a(t)$ with respect to cosmic time. It can be written as a function of the deceleration parameter $q$, a dimensionless measure related to the second order derivative of $a(t)$.  Specifically, the relationship is given by
\begin{equation}
    j \equiv \frac{\frac{d^{3}a}{dt^3}}{aH^3}=(1+z)\frac{dq}{dz} +q(1+2q).
\end{equation}
One of the notable features of the jerk parameter 
$j$ is that it remains constant at $j=1$ throughout cosmic history in the standard $\Lambda$CDM model. This feature makes $j$ a powerful tool for identifying deviations from the $\Lambda$CDM model and also for probing alternative DE models. Just as variations from the canonical EoS value $\omega_{\Lambda}=-1$ suggest departures from the cosmological constant, any deviation of $j$ from unity can indicate new physics beyond the standard model. Specifically, $j(z) > 0$ corresponds to an increase in acceleration, while $j(z) < 0$ indicates a deceleration of cosmic expansion. Due to its simplicity, the jerk formalism offers a practical and effective framework for assessing such deviations. For this scenario, $j(z)$ varies as a function of the redshift $z$ in the form

\begin{eqnarray}	
 j(z) &=& \frac{1}{2} \left[\left\{3 {\theta_1} \ln \left(W\left(\frac{1}{z+1}\right)\right)+3 {\theta_2} W\left(\frac{1}{z+1}\right)^3+1\right\}\right] \nonumber \\
 &\times& \frac{1}{2} \left[\left\{3 {\theta_1} \ln \left(W\left(\frac{1}{z+1}\right)\right)+3 {\theta_2} W\left(\frac{1}{z+1}\right)^3+2\right\}-\frac{3 \left({\theta_1}+3 {\theta_2} W\left(\frac{1}{z+1}\right)^3\right)}{W\left(\frac{1}{z+1}\right)+1}\right].
\end{eqnarray}

To complement the analysis based on deceleration and jerk parameters, we investigate the $Om$ diagnostic~\cite{refom1,refom2}, which is defined as
\begin{equation}
Om(z) \equiv \frac{E^2(z)-1}{(1+z)^3 -1}, \quad E(z) = \frac{H(z)}{H_0},
\end{equation}
where $H_0$ denotes the Hubble constant at the present epoch. The $Om(z)$ diagnostic offers a straightforward yet effective method for distinguishing DE models \cite{refomaam,ddem1,ddem2,ddem3,refomaam25}. In the case of a flat $\Lambda$CDM universe, $Om(z)$ remains constant and equals the current density parameter of DM. Deviations from this constant nature, particularly at low redshifts, indicate a varying EoS $w(z)$, providing evidence for dynamical DE. Moreover, a rising $Om(z)$ curve indicates that the DE component behaves like a phantom field with $w < -1$, while a decreasing $Om(z)$ curve suggests quintessence-like behavior, corresponding to $w > -1$. \\

Finally, by using Eqs.~(\ref{feq}) and (\ref{eq-weffz}) and substituting 
$\dot{H} = -H(1+z) \frac{dH}{dz}$, Eq.~(\ref{aeq}) can be written as

\begin{equation}
\int \frac{dH}{H(z)} = \int \frac{3 \Big(\theta_1 \ln \big(W(\frac{1}{z+1})\big) 
+ \theta_2 W(\frac{1}{z+1})^3 + 1 \Big)}{2 (z+1)} \, dz
\end{equation}

\begin{equation}
\Rightarrow ~ \ln H(z) = \int \frac{3 \Big(\theta_1 \ln \big(W(\frac{1}{z+1})\big) 
+ \theta_2 W(\frac{1}{z+1})^3 + 1 \Big)}{2 (z+1)} \, dz + C,
\end{equation}

\begin{equation}
\Rightarrow ~ H(z) = H_0 \exp \Bigg[ \int_0^z \frac{3 \Big(\theta_1 \ln \big(W(\frac{1}{z'+1})\big) 
+ \theta_2 W(\frac{1}{z'+1})^3 + 1 \Big)}{2 (z'+1)} \, dz' \Bigg],
\end{equation}
where $C$ is the constant of integration and $H_0 = e^C$ is the Hubble constant today (at $z = 0$). Since the integral contains the Lambert $W$ function within a logarithm and a cubic term, it cannot be solved analytically. As a result, $H(z)$ must be numerically evaluated for given values of $\theta_1$, $\theta_2$, and $z$. By jointly examining the evolution of $q(z)$, $j(z)$, and $Om(z)$, we can trace important transitions in cosmic expansion and detect possible departures from the standard $\Lambda$CDM framework.\\

To provide a basis for comparison with our parameter estimations, we also include the standard $\Lambda$CDM scenario in which the cosmic expansion rate takes the form
\begin{equation}
H(z) = H_{0} {\left[{ \Omega_{\Lambda} +\Omega_{m}(1+z)^{3}}\right]}^{\frac{1}{2}},
\end{equation}
where $\Omega_{\Lambda} = (1 - \Omega_{m})$, with $\Omega_{m}$ denotes the current value of the density parameter of DM.

\section{Observational data sets and methodology}\label{sec-data}
To constrain the cosmological parameters of our model, we adopt a Markov Chain Monte Carlo (MCMC) approach that samples the parameter space according to the goodness of fit between theory and observation. The fit is quantified through the chi-square statistic $\chi^{2}$, defined as  \cite{Padilla_2021}
\begin{equation}
    \chi^{2} = \sum_{j} \left[\frac{d_{j}-t_{j}(\theta)}{\sigma_{j}}\right]^{2},
\end{equation}
where $t_{j}(\theta)$ represents the theoretical prediction derived from the model parameters $\theta$, $d_{j}$ denotes the $j$-th observed data point, and $\sigma_{j}$ is the corresponding observational uncertainty. Parameter estimation is executed with the Python package \texttt{emcee} \cite{emcee,emcee1}, which provides an efficient and flexible implementation of the MCMC algorithm, widely employed in the examination of cosmic observational data.\\
 
In this paper, we consider three independent sets of observational data to constrain the parameters of our model.
 
\begin{enumerate}
\item \textbf{Type Ia Supernovae (SNIa):} Type Ia supernovae are regarded as standard candles, since their peak luminosities exhibit remarkably small intrinsic scatter, enabling them to function as trustworthy distance markers. The observed distance modulus is expressed as
\begin{equation}
\mu_{\mathrm{obs}} = m - M,
\end{equation}
where $m$ and $M$ are the apparent and absolute magnitudes, respectively. Within the cosmological framework, the theoretical value of the distance modulus is given by
\begin{equation}
\mu_{\mathrm{th}}(z) =  25 + 5\log_{10} d_L(z) + M_b,
\end{equation}
where $M_{b}$ is the absolute magnitude of SNIa and  $d_L(z)$ denotes the luminosity distance, described as
\begin{equation}
d_L(z) = (1+z) \frac{c}{H_0}\int_0^z \frac{dz'}{E(z')}. \label{dL}
\end{equation}

The Pantheon+ compilation, which includes 1701 Type Ia supernovae, is used in this work \cite{sniaurl} --- an expansion of the earlier Pantheon sample that included 1048 objects. The data set spans the redshift range $0.001<z<2.26$ and represents the light curves of 1550 spectroscopically confirmed supernovae \cite{Scolnic_2022, Brout_2022}. This extensive sample provides one of the most accurate and consistent probes of the late-time expansion of the Universe.\\

\item \textbf{Baryon Acoustic Oscillations (BAO):} The cosmological length scale provided by BAO measurements is used as a standard ruler to measure the expansion history of the Universe. For this work, we use BAO measurements from the SDSS Baryon Oscillation Spectroscopic Survey (BOSS) \cite{PhysRevD.103.083533},  the 6dF Galaxy Survey (6dFGS) \cite{Beutler_2011}, and the first-year release of DESI \cite{desicollaboration2024desi2024vicosmological}. From these data sets, we derive several distance measures --- the transverse comoving distance, the comoving horizon distance, and the volume-averaged distance that combines both transverse and radial information, defined as
\begin{align}
    D_M&=\frac{d_L}{1+z},\\
    D_H&=\frac{c}{H(z)}, \\   
    D_V&=\left[\frac{d_L}{1+z}\right]^{2/3}\left[\frac{cz}{H(z)}\right]^{1/3},
\end{align}
respectively, where $d_L$ denotes the luminosity distance (see Eq.~\eqref{dL}). When divided by the sound horizon at the drag epoch, $r_d$, the dimensionless quantities $D_M/r_d$, $D_H/r_d$  and $D_V/r_d$ are obtained, as reported in Table \ref{tab:baodata}. These ratios are widely used as observables in BAO studies, offering stringent constraints on cosmological parameters and allowing tests of deviations from the standard $\Lambda$CDM model.

    \begin{table}[htbp]
		\centering
		\caption{BAO data measurements from the SDSS survey \cite{PhysRevD.103.083533}, the 6dFGS survey \cite{Beutler_2011}, and DESI DR1 \cite{desicollaboration2024desi2024vicosmological}. The data set lists the effective redshifts ($ z_\text{\rm{eff}} $) and the corresponding values for the ratios $ D_{\rm H}/r_{\rm d} $, $ D_{\rm M}/r_{\rm d} $  and $ D_{\rm V}/r_{\rm d} $. The BAO data used in this study is taken from Ref. \cite{Luongo:2024zhc}.}
			
		\begin{tabular}{lcccc}
			\hline
			Survey          & $z_\text{\rm{eff}}$ & $D_{\rm M}/r_{\rm d}$ & $D_{\rm H}/r_{\rm d}$ & $D_{\rm V}/r_{\rm d}$\\
			
			\hline		
            SDSS MGS 		& $0.15$	& & & $4.51\pm0.14$ \\
			SDSS DR12		& $0.38$    & $10.27\pm0.15$ & $24.89\pm 0.58$ & \\
			SDSS DR16 ELG	& $0.85$	& $19.50\pm1.00$ & $19.60\pm2.10$ & \\
			SDSS DR12		& $0.51$    & $13.38\pm0.18$ & $22.43\pm 0.48$ & \\
			SDSS DR16 QSO	& $1.48$    & $30.21\pm0.79$ & $13.23\pm0.47$ & \\
			SDSS DR16 LRG	& $0.70$    & $17.65\pm0.30$ & $19.78\pm0.46$ & \\
			SDSS DR16 Ly$\alpha$-Ly$\alpha$& $2.33$     & $37.60\pm1.90$& $8.93\pm0.28$ & \\
			SDSS DR16 Ly$\alpha$-QSO	& $2.33$ & $37.30\pm1.70$& $9.08\pm0.34$ & \\
			
            \hline
			6dFGS		    & $0.106$	& & & $2.98\pm0.13$ \\
			\hline
            DESI BGS 		& $0.30$    & & & $7.93\pm0.15$ \\
			DESI LRG1	 	& $0.51$ 	& $13.62\pm0.25$ & $20.98\pm0.61$ & \\
			DESI ELG        & $1.32$    & $27.79\pm0.69$ & $13.82\pm0.42$ & \\
			DESI LRG2       & $0.71$ 	& $16.85\pm0.32$ & $20.08\pm0.60 $& \\
			DESI Ly$\alpha$-QSO & $2.33$ 	& $39.71\pm0.94$ & $8.52\pm0.17$ & \\
			DESI LRG+ELG    & $0.93$    & $21.71\pm0.28$ & $17.88\pm0.35$ & \\			
			DESI QSO		& $1.49$	& & & $26.07\pm0.67$ \\
			\hline
		\end{tabular}
		\label{tab:baodata}
	\end{table}

\item \textbf{Observational $H(z)$ Data (OHD) or Cosmic Chronometers (CC):} By measuring the Hubble parameter $H(z)$ at various redshifts, cosmic chronometers offer a straightforward, model-independent way to calculate the Universe's expansion rate by measuring the differential age evolution of passively evolving galaxies. For our analysis, we use the most recent compilation from Ref. \cite{Favale_2023}, which includes 32 $H(z)$ measurements covering the redshift range $0.07<z<1.965$. The data set accounts for both systematic and statistical uncertainties, making it a valuable resource for constraining cosmological parameters in combination with other probes.
\end{enumerate}

To obtain the final, comprehensive, cosmological constraints, we construct a joint likelihood function 
$\mathcal{L}$  by combining the independent likelihoods of the individual observational data sets. The total likelihood is expressed as
\begin{equation}
\mathcal{L} = \mathcal{L}_{\text{SNIa}} \times \mathcal{L}_{\text{BAO}} \times \mathcal{L}_{\text{OHD}},
\end{equation}
where each term corresponds to the likelihood associated with SNIa, BAO, and OHD data sets, respectively. Since these data sets are statistically independent, their combined likelihood can equivalently be represented in terms of the total chi-square function
\begin{equation}\label{eqtotalchi2m}
\chi^2_{\text{tot}} = \chi^2_{\text{SNIa}} + \chi^2_{\text{BAO}} + \chi^2_{\text{OHD}}.
\end{equation}
This total $\chi^2_{\text{tot}}$ serves as the quantity minimized in the MCMC analysis to determine the best-fit cosmological parameters. The  maximum likelihood, $ \mathcal{L}_{max}$, corresponds to the minimum chi-squared value ($\chi^{2}_{min}$), such that 
\begin{equation}
\chi^{2}_{min}= - 2\rm{ln}(\mathcal{L}_{max}).    
\end{equation}

Furthermore, we shall test the statistical preference of our model against the $\Lambda$CDM model by computing the Akaike Information Criterion (AIC) and the Bayesian Information Criterion (BIC), and analyzing their deviations from their corresponding minimum values.\\

The AIC was developed by the Japanese statistician Hirotugu Akaike \cite{97} to assist in selecting the best model without overfitting. In simpler terms, AIC prefers models that fit the data well, but inflicts penalty for each additional parameter. The AIC for a model is computed as
\begin{equation}
\text{AIC} = -2 \ln(\mathcal{L_{\rm{max}}})+2k,
\end{equation}
where $\mathcal{L_{\rm{max}}}$ is the maximum likelihood and $k$ is the number of free parameters. To facilitate objective comparisons between different cosmological models, those with lower AIC values are preferred. In particular, $0 < \Delta \rm{AIC} < 2$ indicates that the test and reference models are statistically similar, $4 < \Delta \rm{AIC} < 7$ indicates a significant preference for the reference model, while $\Delta \rm{AIC} > 10$ suggests a substantial preference for the reference model. In our case, the Lambert $W$ model is the test model, while the $\Lambda$CDM model serves as the reference model.\\ 

The BIC was introduced by Schwarz \cite{98} and is calculated as
\begin{equation}
\text{BIC} = - 2 \ln(\mathcal{L_{\rm{max}}}) + k \ln n,
\end{equation}
where $n$ is the number of data points. It arises from approximations of the evidence ratios of models, known as the Bayes factor \cite{Jeffreys1,Kass1,Burnham1}. The data points are assumed to be independent and identically distributed in the BIC. This may or may not be valid depending on the data set under consideration, for example, it may be well-suited for supernova luminosity–distance data but not likely to be good for cosmic microwave anisotropy data \cite{Liddle2007}. The strength of evidence is described as follows \cite{Kass1} --- $0 < \Delta \rm{BIC} < 2$ indicates weak or statistically equal preference for the models in question, $2 < \Delta \rm{BIC} < 6$ indicates moderate evidence for the reference model, while $6 < \Delta \rm{BIC} < 10$ and $\Delta \rm{BIC} > 10$, respectively, suggest strong and decisive evidence for the reference model \cite{Anag1,Bhatt1}.\\

Both AIC and BIC attempt to resolve the problem of overfitting by penalizing the number of free parameters of a model, however, the penalty imposed by BIC is more stringent when compared to that imposed by AIC.

\section{Results and Discussion}\label{sec-results}
\begin{figure}[htbp]
\begin{center}			
\includegraphics[width=15cm,height=15cm]{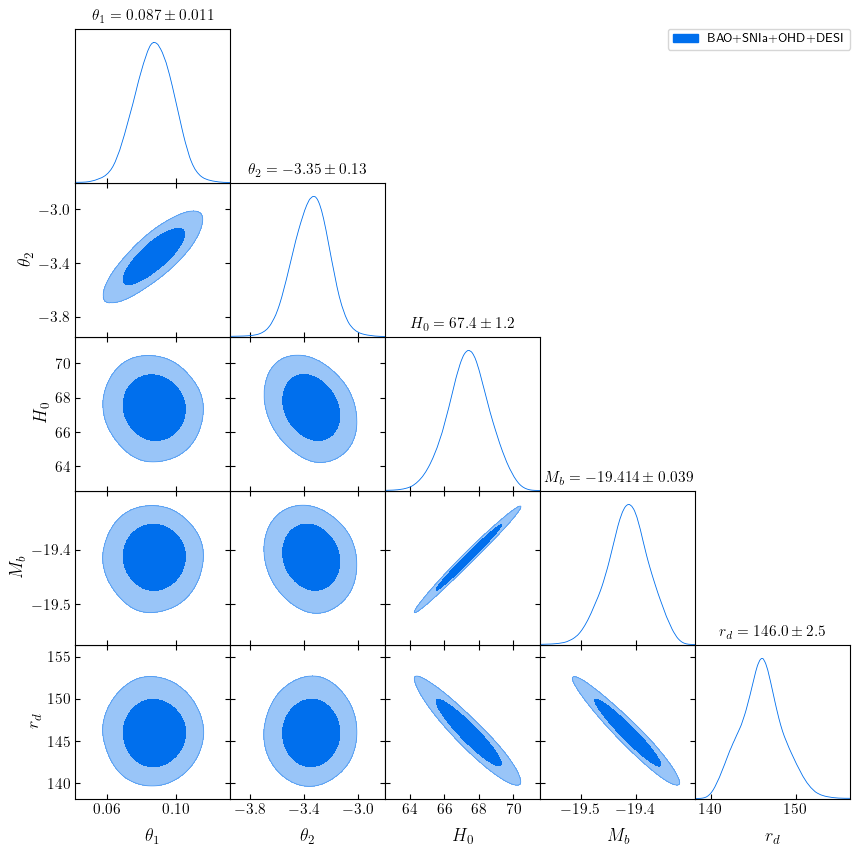}
\caption{Triangle plot showing the $1\sigma$ and $2\sigma$ confidence contours, along with the one-dimensional marginalized posterior distributions of the Lambert $W$ model parameters, obtained from the combined SNIa+BAO+OHD data set.}\label{figc1}
\end{center}
\end{figure}

\begin{figure}[ht]
\begin{center}			
\includegraphics[width=15cm,height=15cm]{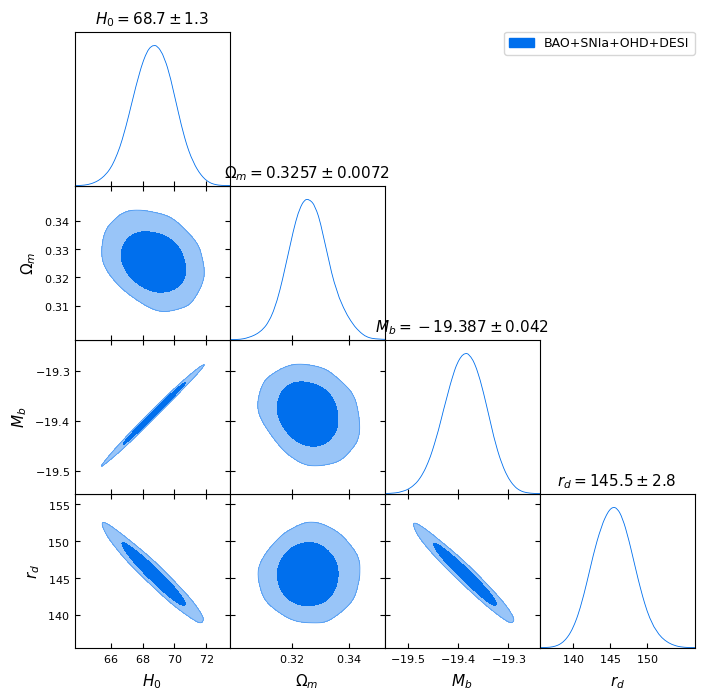}
\caption{ Triangle plot showing the $1\sigma$ and $2\sigma$ confidence contours, along with the one-dimensional marginalized posterior distributions of the $\Lambda$CDM model parameters, obtained from the combined SNIa+BAO+OHD data set.}\label{figc2}
\end{center}
\end{figure}

We employed the popular Markov Chain Monte Carlo (MCMC) statistical technique to obtain the best-fit values for the model parameters. We used the \texttt{emcee} Python package, created by Foreman-Mackey et al.~\cite{emcee}, to estimate the model parameters. For the MCMC analysis on the combined SNIa+BAO+OHD data sets, we used 20000 iterations (steps) and 64 random chains (walkers). We obtained initial estimates for the model parameters by minimizing Eq. (\ref{eqtotalchi2m}) using the Python package \texttt{scipy.optimize}. Figs.~\ref{figc1} and \ref{figc2} display the 1$\sigma$ and $2\sigma$ confidence intervals for the key cosmological parameters, in the Lambert $W$ and $\Lambda$CDM models, respectively. \\

Table~\ref{tab:results} summarizes the constraints on the parameters obtained for the Lambert $W$ (column III) and $\Lambda$CDM (column IV) models, while also highlighting the key limits imposed on each parameter. The best-fit values of the free parameters $\theta_1$ and $\theta_2$ in the Lambert $W$ model are obtained as $\theta_1=0.087$ and $\theta_2=-3.35$. These values show significant agreement with the theoretically estimated values $\theta_1 \approx \frac{1}{7}$ and $\theta_2 \approx -\frac{16}{5}$ that the authors in Ref. \cite{Saha1} obtained using information from the observational analysis performed with only the Hubble parameter data. Furthermore, the Lambert $W$ EoS yields a Hubble constant value $H_{0} = 67.4 \pm 1.2~\mathrm{km,s^{-1},Mpc^{-1}}$, which is in excellent agreement with the Planck estimate, $H_0 = 67.36 \pm 0.54~\mathrm{km,s^{-1},Mpc^{-1}}$ \cite{acc5}. However, it deviates significantly from the value reported by the SH0ES collaboration, $H_0 = 73.04 \pm 1.04~\mathrm{km,s^{-1},Mpc^{-1}}$ \cite{refh02}, thereby reflecting the persistent Hubble tension. In addition, current measurements of the sound horizon at the drag epoch yield $r_d = 146\pm 2.5$~Mpc, consistent with recent BAO observations \cite{Ham1,refhoalle1,refhoalle2}. Notably, the value of the Hubble constant reported in Ref.~\cite{refRevjcap}, $H_0 = 67.5^{+1.3}_{-1.6}~\mathrm{km,s^{-1},Mpc^{-1}}$, and obtained using a parameterized Hubble-function framework constrained by a similar data set, is consistent with our estimate within the quoted uncertainties. Such agreement between independent analyses lends further credibility to the present results and highlights their robustness. However, since the underlying model assumptions differ from those adopted in our analysis, we restrict the comparison to a qualitative consistency check rather than a detailed quantitative analysis. The information criteria, AIC and BIC, and their corresponding relative differences for the two models are summarized in Table \ref{tab:AIC}. The $\Delta$AIC values suggest that, although the Lambert $W$ model is statistically similar to the standard $\Lambda$CDM model, it does not represent the optimal choice due to a larger set of parameters. However, the $\Delta$BIC values show strong evidence against the Lambert $W$ model. This is probably due to the fact that BIC has a more stringent penalty term for the number of free parameters. So, we cannot be very sure whether the Lambert $W$ model is truly inferior compared to the $\Lambda$CDM model.\\ 

\begin{figure}[ht]
	\begin{center}
		   \includegraphics[width=12cm,height=8cm]{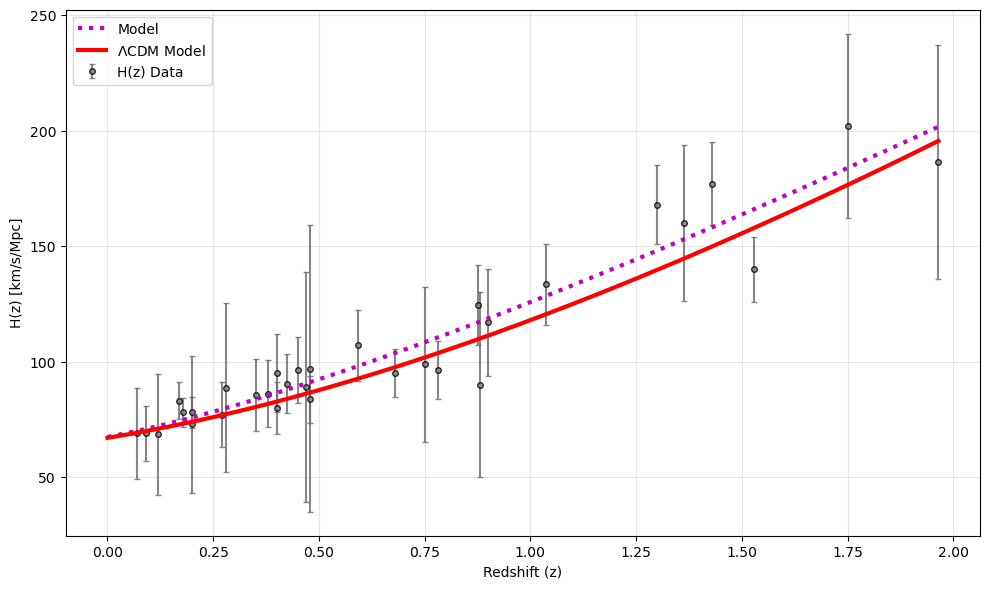}          
	\caption{The best-fit evolution of the Hubble parameter $H(z)$ for the Lambert $W$ model, obtained from the combined SNIa+BAO+OHD data set. The solid curve represents the best-fit $H(z)$ prediction of the Lambert $W$ model, while the observational data points with error bars correspond to the compiled measurements. For comparison, the predicted 
$H(z)$ evolution from the standard $\Lambda$CDM model is also shown.}\label{fighz}
	\end{center}
\end{figure}

\begin{figure}[ht]
		\begin{center}
		\includegraphics[width=12cm,height=8cm]{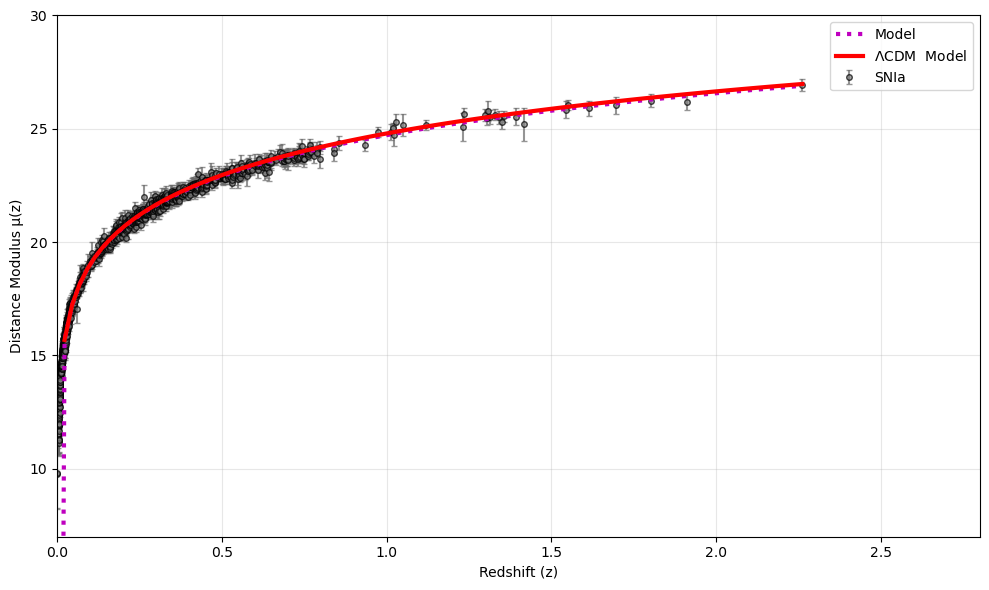}
		\caption {The best-fit evolution of the distance modulus $\mu(z)$ for the Lambert $W$ model, derived from the combined SNIa+BAO+OHD data set. The observational data points with corresponding uncertainties are shown for comparison, along with the predicted $\mu(z)$ evolution from the standard $\Lambda$CDM model.}\label{figmuz}	
	\end{center}	
	\end{figure}

Figures~\ref{fighz} and~\ref{figmuz} illustrate the best-fit evolutions of the Hubble parameter $H(z)$ and the distance modulus $\mu(z)$, respectively, as obtained with the combined SNIa+BAO+OHD data set. For comparison, the corresponding predictions of the $\Lambda$CDM model are also included in the same figures. These results suggest that the evolutions of $H(z)$ and $\mu(z)$ in the Lambert $W$ model are not always consistent with those in the $\Lambda$CDM model, particularly at higher redshifts, however, at low redshifts ($z \lesssim 0.5$), their behavior remains broadly compatible with both the observational data and the predictions of the standard $\Lambda$CDM model.\\

 \renewcommand{\arraystretch}{1.5} 
\begin{table}[htbp]
		\centering
		\caption{Constraints on the parameters with prior ranges and $95\%$ credible limits of the model.}
		\begin{tabular}{c c c c}
			\hline \hline
			Parameter & Prior & Lambert $W$ & ~~$\Lambda$CDM\\
			\hline \hline			
			{\boldmath$\theta_{1}             $} & $(0,0.5)$ & $0.087^{+0.011}_{-0.011}$ & -\\
            \hline
			{\boldmath$\theta_{2}             $} & $(-4, -2)$ & $-3.35^{+0.13}_{-0.13}$ & -\\
            \hline
		    {\boldmath$H_{0}       $} & $(50, 100)$ & $67.4^{+1.2}_{-1.2}$ & ~~$68.7^{+1.3}_{-1.3}$\\
            \hline
			{\boldmath$M_b            $} & $(-21, -18)$ & ~~$-19.414^{+0.039}_{-0.039}$ & ~~$-19.387^{+0.042}_{-0.042}$\\
			\hline
			{\boldmath$r_d            $} & $(140, 160)$ & $146.0^{+2.5}_{-2.5} $ & ~~$145.5^{+2.8}_{-2.8} $\\
			\hline
			{\boldmath$\Omega_{m}            $} & $(0,1)$ & - & ~~$0.3257^{+0.0072}_{-0.0072} $\\
			\hline \hline
		\end{tabular}
		\label{tab:results}
	\end{table}
\begin{table}
\caption{\small The $\chi^2_\text{tot,min}$ and AIC, BIC together with their corresponding differences for the $\Lambda$CDM and Lambert $W$ models.}
	\begin{center}
		\begin{tabular}{c c c c c c}
			\hline \hline
			Model  & $\chi^2_\text{tot,min}$ &  	AIC & $ \Delta \rm{AIC}$  &  	BIC & $ \Delta \rm{BIC}$\\			
			\hline \hline				
			$\Lambda$CDM	& 914.24& 922.24 &  0 	& 944.14	& 0  \\			
			\hline				
			Lambert $W$	&  913.65	& 923.65 &  1.41 & 951.02	& 6.88\\
			\hline \hline			
		\end{tabular}
		\label{tab:AIC}
	\end{center}
\end{table}	

\begin{figure}
\begin{center}				
\includegraphics[width=12cm,height=8cm]{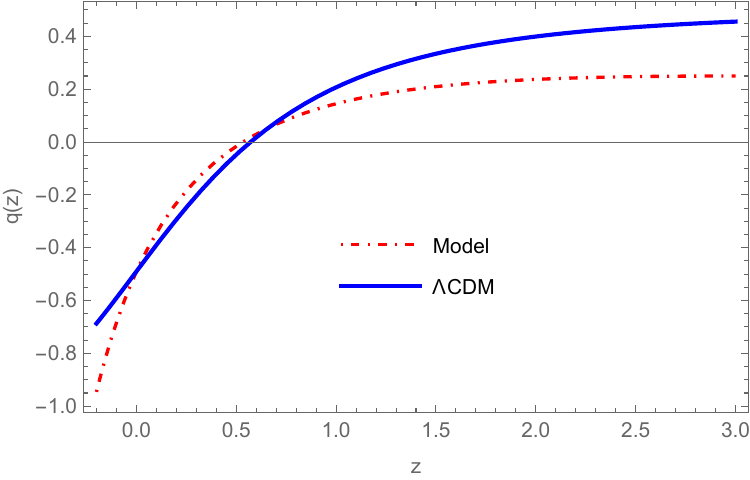}
\caption{The best-fit curve of the deceleration parameter $q$ as a function of the redshift $z$ in the Lambert $W$ and $\Lambda$CDM models obtained from the combined SNIa+BAO+OHD data set. The horizontal line indicates the point where $q(z)=0$.}\label{figqz}
\end{center}
\end{figure}
\begin{figure}
\begin{center}				
\includegraphics[width=12cm,height=8cm]{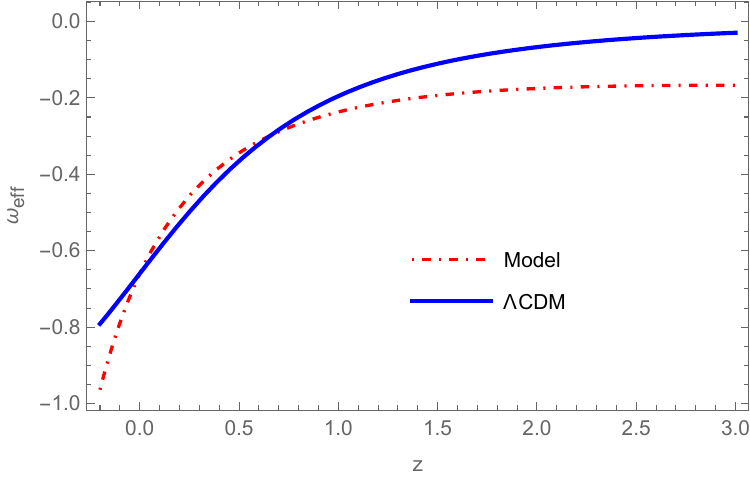}
\caption{The best-fit curve of the effective EoS parameter $\omega_{\rm{eff}}$ as a function of the redshift $z$ in the Lambert $W$ and $\Lambda$CDM models, obtained from the combined SNIa+BAO+OHD data set.}\label{figweffz}
\end{center}
\end{figure}
\begin{figure}
\begin{center}				
\includegraphics[width=12cm,height=8cm]{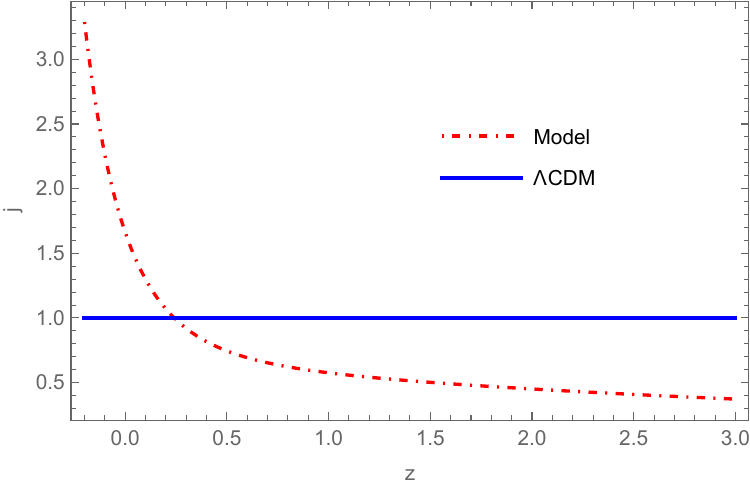}
\caption{The best-fit curve of the jerk parameter $j$ as a function of the redshift $z$ in the Lambert $W$ and $\Lambda$CDM models, obtained from the combined SNIa+BAO+OHD data set.}\label{figjz}
\end{center}
\end{figure}
\begin{figure}
\begin{center}				
\includegraphics[width=12cm,height=8cm]{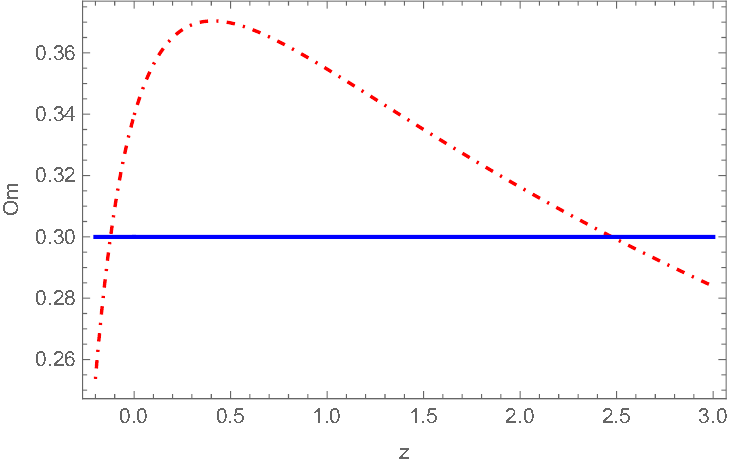}
\caption{The best-fit curve of the $Om(z)$ diagnostic in the Lambert $W$ and $\Lambda$CDM models, obtained from the combined SNIa+BAO+OHD data set.}\label{figomz}
\end{center}
\end{figure}

Finally, to examine the evolutionary behavior of our model, we plotted the deceleration parameter $q(z)$, the effective EoS $\omega_{\rm{eff}}(z)$, the jerk parameter $j(z)$,  and the $Om(z)$ diagnostic against the redshift $z$, and presented them in Figs. \ref{figqz}, \ref{figweffz}, \ref{figjz}, and \ref{figomz}, in this order. For comparison, the evolutionary behaviors of the corresponding parameters in the $\Lambda$CDM model are also presented. Fig. \ref{figqz} reveals a clear transition from a decelerated universe to an accelerated one. Interestingly, the estimated transition redshift $z_{t}\approx 0.6$ obtained from the combined SNIa+BAO+OHD data set align with previous studies \cite{refzt1,refzt2,refzt3,refzt4,refzt5,refzt6,refzt7,refzt8}. The evolution of the Lambert $W$ EoS in Fig.~\ref{figweffz} shows that $\omega_{\rm{eff}}(z)$ approaches zero at high redshifts, for the best-fit values of $\theta_1=0.087$ and $\theta_2=-3.35$ obtained from our analysis, thus effectively behaving like DM in the early Universe. At lower redshifts, $\omega_{\rm{eff}}(z)$ transitions into the quintessence regime, with a current value of approximately $\omega_{{\rm{eff}},0} = -0.75$. This shows that the functional form of $\omega_{\rm{eff}}(z)$, as defined in Eq.~(\ref{eq-weffz}), can seamlessly describe both the early matter-dominated era and the late-time DE-dominated era. The behavior of the jerk parameter $j(z)$ in Fig. \ref{figjz} shows clear deviations from the constant value $j=1$ predicted by the standard $\Lambda$CDM model. This observation incorporates the dynamical nature of the Lambert $W$ model. Finally, the evolution of the $Om(z)$ diagnostic in Fig.~\ref{figomz} shows a steady increase from higher redshifts to $z \sim 0.5$, gains a maximum, and then falls steeply at low redshifts, and this behavior continues into the future. Thus, the $Om(z)$ diagnostic also reveals dynamical nature of the Lambert $W$ EoS.

\section{Summary and conclusions}\label{sec-cons}
This work dealt with a comprehensive observational data analysis of the Lambert $W$ equation of state (EoS), first proposed by Saha and Bamba in 2019 \cite{Saha1}. Our particular interest was to constrain the two free parameters, $\theta_1$ and $\theta_2$, that appear in the Lambert $W$ EoS. We used latest data obtained from the Pantheon$+$ compilation of Type Ia Supernovae (SNIa), Baryon Acoustic Oscillations (BAO) measurements from the SDSS Baryon Oscillation Spectroscopic Survey (BOSS), the 6dF Galaxy Survey (6dFGS), and the first-year release of the Dark Energy Spectroscopic Instrument (DESI), and observational $H(z)$ (OHD) data, otherwise known as Cosmic Chronometers (CC), to perform a joint analysis on the $\Lambda$CDM and the Lambert $W$ models. We obtained the best-fit values $\theta_1=0.087$ and $\theta_2=-3.35$, which agree well with the theoretically estimated values \cite{Saha1}. The best-fit values for the Hubble constant $H_0$, the peak absolute magnitude $M_b$ of SNIa, and the sound horizon $r_d$ at the drag epoch are obtained, respectively, as $67.4$ km/s/Mpc, $-19.414$, and $146.0$ Mpc, consistent with recent observations. As discussed in the previous section, the result reported in Ref.~\cite{refRevjcap}, $H_0 = 67.5^{+1.3}_{-1.6}~\mathrm{km\,s^{-1}\,Mpc^{-1}}$, obtained within a parameterized Hubble-function framework using a similar data set, is consistent with our estimate, providing an independent cross-check and further supporting the robustness of our findings. Our analysis indicates that the inferred Hubble constant is in agreement with the Planck estimate ($H_0 = 67.36 \pm 0.54~\mathrm{km\,s^{-1}\,Mpc^{-1}}$) \cite{acc5}, but shows significant tension with the SH0ES value ($H_0 = 73.04 \pm 1.04~\mathrm{km\,s^{-1}\,Mpc^{-1}}$) \cite{refh02}. Moreover, this Hubble tension may also persist in the SNIa+BAO+OHD data set when the Lambert $W$ equation of state in Eq.~(\ref{eq-weffz}) is adopted. In addition, the best-fit curves for the Hubble parameter $H(z)$ and the distance modulus $\mu(z)$ in the Lambert $W$ model are illustrated in Figs. \ref{fighz} and~\ref{figmuz} respectively. They show deviations from those of the $\Lambda$CDM model at higher redshifts, however, at lower redshifts, their evolutions show compatibility with those obtained in the $\Lambda$CDM model.\\ 

We also tested our model with the information criteria, particularly the Akaike Information Criterion (AIC) and the Bayesian Information Criterion (BIC). The values of $\Delta AIC$ revealed that the Lambert $W$ model is statistically similar to the $\Lambda$CDM model, however, the former cannot be considered superior due to a larger set of parameters. However, the values of $\Delta$BIC revealed strong evidence against the Lambert $W$ model, probably because the BIC penalizes the number of free parameters more strongly compared to the AIC. Thus, it may be an overstatement that the Lambert $W$ model is inferior, if not superior, to the concordance $\Lambda$CDM model.\\

The evolutions of the deceleration parameter $q(z)$, the jerk parameter $j(z)$, the effective EoS $\omega_{\rm{\rm{eff}}}(z)$, and the $Om(z)$ diagnostic are illustrated in Figs. \ref{figqz} through \ref{figomz} respectively. The deceleration parameter shows a transition from the matter-dominated decelerated era to the late-time cosmic acceleration era at $z_t \approx 0.6$. The evolution of the effective EoS shows that the Lambert $W$ model mimics a pressureless DM-dominated era at higher redshifts and decreases uniformly until it attains the value $\omega_{\rm{eff}}=-1$ in the future. The present value of the effective EoS is approximately $0.75$, consistent with that obtained in the $\Lambda$CDM model. Thus, Lambert $W$ EoS can be regarded as a potential candidate for a single dark-dark fluid. Finally, the variations in the jerk parameter and the $Om(z)$ diagnostic reveal the dynamical nature of the Lambert $W$ EoS. These observations require a deeper, more comprehensive analysis of this single dark-fluid model. \\

Overall, our results demonstrate that the Lambert $W$ EoS is observationally consistent, physically viable, and offers a dynamically rich description of the cosmic expansion history.  To fully assess the viability of the Lambert $W$ model, further aspects need to be investigated. The present work can be extended by incorporating a broader set of observational data sets, including BAO measurements from DESI DR2, SNIa data from DESY5, and Hubble parameter constraints from gravitational-wave standard sirens of neutron star binaries. Confrontation with high-precision cosmological probes, particularly $\sigma_8$ measurements, could provide a stringent test of the model. These data sets are essential for refining parameter estimates and judging the consistency of the model across cosmic time. Ongoing efforts focus on integrating these data sets. These improved and updated results shall be presented in future studies.

\begin{acknowledgments}
The author S.S. thanks Supriya Pan for very useful and enlightening discussions on the information criteria and their importance in cosmological model selection. All the authors thank the reviewer for a careful reading of the manuscript and for the valuable comments and suggestions, which have helped improve its quality and presentation significantly. 
\end{acknowledgments}


\end{document}